\documentclass[12pt]{article}
\begin{document}
\title{\bf Dickman polylogarithms and their constants}

\author{David Broadhurst\thanks{Department of Physics and Astronomy,
The Open University, Milton Keynes, MK7 6AA, United Kingdom,
{\tt D.Broadhurst@open.ac.uk}}}

\date{\today}
\maketitle

\abstract{The Dickman function $F(\alpha)$ gives the asymptotic
probability that a large integer $N$ has no prime divisor exceeding
$N^\alpha$. It is given by a finite sum of generalized
polylogarithms defined by the exquisite recursion
$L_k(\alpha)=-\int_\alpha^{1/k}{\rm d}x
L_{k-1}(x/(1-x))/x$ with $L_0(\alpha)=1$. The behaviour of these
Dickman polylogarithms as $\alpha\to0$ defines an intriguing
series of constants, $C_k$. I conjecture that $\exp(\gamma z)/\Gamma(1-z)$
is the generating function for $\sum_{k\ge0}C_k z^k$.
I obtain high-precision evaluations of $F(1/k)$, for integers $k<11$,
and compare the Dickman problem with problems in
condensed matter physics and quantum field theory.}

\setlength{\parskip}{2mm}

\section{Introduction}

The fraction of positive integers less than $N$
with no prime divisor greater than $N^\alpha$ tends to a finite limit
$F(\alpha)$ as $N\to\infty$. Clearly $F(\alpha)=1$ for $\alpha\ge1$.
For $0<\alpha<1$,
the Dickman~\cite{Di} function $F(\alpha)$ satisfies
the remarkable differential equation~\cite{Ch,Rama,Br,HT}
\begin{equation}
F^\prime(\alpha)=\frac{1}{\alpha}F\left(\frac{\alpha}{1-\alpha}\right)
\label{F1}\end{equation}
and may be computed as a finite sum of terms
\begin{equation}
F(\alpha)=\sum_{k=0}^{K(\alpha)}L_k(\alpha)
\label{F}\end{equation}
where $L_0(\alpha)=1$, $K(\alpha)$ is the largest integer
such that $\alpha K(\alpha)<1$, and the recursion
\begin{equation}
L_k(t)=-\int_t^{\frac1k}L_{k-1}\left(\frac{x}{1-x}\right)\frac{{\rm d}x}{x}
\label{L}\end{equation}
defines the Dickman polylogarithm of weight $k>0$ as an iterated integral.
I shall prove two theorems that allow me to obtain at least 100 digits
of $L_k(t)$ for all weights $k<10$ and hence shall conjecture the
generating function for the Dickman constants
\begin{equation}
C_k=\lim_{t\to0}\left(L_k(t)-\sum_{j=1}^k
\frac{C_{k-j}\log^j(t)}{j!}\right)
\label{C}\end{equation}
where $C_0=1$ and $C_k$ is determined by
the boundary condition $L_k(\frac1k)=0$ for $k>0$.

\section{Dickman trilogs from standard trilogs}

Since $L_1(t)=\log(t)$, we have $C_1=0$ and $F(1/2)=1-\log(2)$.
Moreover, we easily obtain~\cite{Cham}
\begin{equation}
L_2(t)=-\int_t^{\frac12}\log\left(\frac{x}{1-x}\right)\frac{{\rm d}x}{x}
={\rm Li}_2(t)+\frac12\log^2(t)-\frac{\pi^2}{12}
\label{L2}\end{equation}
where ${\rm Li}_k(t)=\sum_{n>0}t^n/n^k$ is the standard
polylogarithm of weight $k$ and $C_2=-\frac{\pi^2}{12}$
ensures that $L_2(\frac12)=0$. Thus
\begin{eqnarray}
F(1/3)&=&1-\log(3)+
\sum_{n=1}^\infty\frac{1}{n^2 3^n}+\frac12\log^2(3)-\frac{\pi^2}{12}
\label{F13}\\&\approx&
0.048608388291131566907183039343407421354329580478141
\label{F31n}\end{eqnarray}
may be computed with great facility.

After considerable effort, I shall prove in Section~5 that
\begin{eqnarray}
L_3(t)&=&{\rm Li}_3(2t-1)-{\rm Li}_3(1-2t)-{\rm Li}_3(t)-{\rm Li}_3(2-1/t)
\nonumber\\
&&{}+\log\left(\frac{t}{1-2t}\right)L_2(t)
+\frac{\pi^2}{6}\log(t)-\frac16\log^3(t)+\frac{17}{12}\,\zeta(3)
\label{L3}\end{eqnarray}
and hence obtain the Dickman constant
\begin{equation}
C_3=\lim_{t\to0}\left
(L_3(t)+\frac{\pi^2}{12}\log(t)-\frac16\log^3(t)\right)
=-\frac13\,\zeta(3).
\label{C3}\end{equation}

It follows from~(\ref{F},\ref{L2},\ref{L3}) that $F(\alpha)$
reduces to standard polylogs for $\alpha\ge\frac14$. In particular,
the asymptotic probability that an integer $N$ has no
prime divisor greater than $N^\frac14$ is
\begin{eqnarray}
F(1/4)&=&1-2\log(2)
+{\rm Li}_2\left(\frac14\right)+2\log^2(2)-\frac{\pi^2}{12}
\nonumber\\&&{}
-{\rm Li}_3\left(\frac14\right)
-{\rm Li}_2\left(\frac14\right)\log(2)
-\frac23\log^3(2)+\frac{13}{24}\,\zeta(3)
\label{F14}\\&\approx&
0.0049109256477608323527391509236151860324842974176929
\label{F14n}\end{eqnarray}
with 800 good digits obtainable in less than 10 milliseconds.

\section{Conjecture for the Dickman constants}

I shall show, in later sections, how to evaluate $F(\alpha)$,
for $\frac14>\alpha\ge\frac{1}{10}$,
and the intriguing constants $C_k$, for $3<k<10$,
by one-dimensional numerical quadrature, at precisions
well in excess of 100 decimal digits. This
is amply sufficient to obtain very reliable conjectures
for the analytic form of $C_k$.
The complexity of these calculations is in marked contrast to the
final simplicity of the conjectured results. For example,
\begin{eqnarray}
C_4&=&\int_0^{\frac12}
\left(
\log\left(\frac{x}{1+2x}\right)\,{\rm Li}_2(x)
+\frac12\log^2(x)\,{\rm Li}_1(-2x)
\right)\frac{{\rm d}x}{x(1+x)}
+3\,{\rm Li}_4\left(\frac12\right)
\nonumber\\&&{}
-\frac38\,{\rm Li}_4\left(\frac14\right)
-\frac{3\log(2)}{4}\,{\rm Li}_3\left(\frac14\right)
+\frac{\pi^2-9\log^2(2)}{12}\,{\rm Li}_2\left(\frac14\right)
+\frac{21\log(2)\zeta(3)}{8}
\nonumber\\&&{}
+\frac{\pi^2\log^2(2)}{24}
-\frac{\pi^2\log(2)\log(3)}{6}
+\frac{\log^3(2)\log(3)}{2}
-\frac{5\log^4(2)}{8}
\label{C4}
\end{eqnarray}
has a very simple conjectural evaluation as $\frac1{10}C_2^2$,
which holds at 800-digit precision.
The complexity of evaluation~(\ref{C4})
results from delicate integration by parts, so as to
remove divergent terms at the lower limit of
integration as $t\to0$ in~(\ref{L}) and leave a finite integral
whose integrand contains no trilogs.

The constant $C_8$ is of particular interest, since that is
the first case where a multiple zeta
value (MZV)~\cite{DZ,BBB1,BBB2,BBB3,BBV}
might have occurred.
However, after far more intricate numerical quadratures, the LLL~\cite{LLL}
algorithm implemented by the Pari-GP~\cite{PARI} procedure {\tt lindep}
gave the conjectured evaluation
\begin{equation}
C_8\;\stackrel{?}{=}\;\frac{\zeta(5)\zeta(3)}{15}
-\frac{\left[\zeta(3)\pi\right]^2}{216}-\frac{67\pi^8}{29030400}
\label{C8}\end{equation}
which is free of the irreducible MZV of lowest weight,
namely $\zeta(5,3)=\sum_{m>n>0}1/(m^5n^3)$.
After noticing that $-67$ is the first non-unit integer
in the sequence $1,\,-1,\,1,\,-1,\,-67,\,-1,$
given by Neil Sloane~\cite{OEIS} in OEIS entry A008991, for
numerators of coefficients in the expansion of
\begin{equation}
\sqrt{\frac{\sin(x)}{x}}=
1 - \frac{x^2}{12} + \frac{x^4}{1440}
- \frac{x^6}{24192} - \frac{67x^8}{29030400}
- \frac{x^{10}}{5677056} +{\rm O}\left(x^{12}\right),
\label{NS}\end{equation}
I was led to conjecture the wonderfully compact
generating function
\begin{equation}
G(z)\;\equiv\;\frac{\exp(\gamma z)}{\Gamma(1-z)}
\;\stackrel{?}{=}\;\sum_{k=0}^\infty C_k z^k
\label{G}\end{equation}
where $\gamma$ is Euler's constant. The equivalent, yet
more illuminating, formula
\begin{equation}
G(z) = \sqrt{\frac{\sin(\pi z)}{\pi z}}\exp\left(
-\sum_{n>0}\frac{\zeta(2n+1)}{2n+1}z^{2n+1}\right)
\label{G1}\end{equation}
neatly accounts for Sloane's tell-tale integer $67$ in~(\ref{C8})
and leads to the conjecture
\begin{equation}
C_9\;\stackrel{?}{=}\;
-\frac{\zeta(9)}{9}
+\frac{\pi^2\zeta(7)}{84}
-\frac{\pi^4\zeta(5)}{7200}
+\frac{\pi^6\zeta(3)}{72576}
-\frac{\left[\zeta(3)\right]^3}{162}
\label{C9}\end{equation}
which was tested by numerical quadrature of a product of
Dickman tetralogarithms, thanks to the following theorem.

\section{Integration by parts}

I define an auxiliary family of functions by the recursion
\begin{equation}
f_{n+1}(t) = -\int_t^\frac1{n+1}\frac{f_n(x)\,{\rm d}x}{x(1-n x)}
\label{f}\end{equation}
with $f_0(t)=1$. Then $f_1(t)=\log(t)=L_1(t)$ and the dilogarithm
\begin{equation}
f_2(t)=-\int_t^\frac12\frac{\log(x)\,{\rm d}x}{x(1-x)}
=\log\left(\frac{t}{1-t}\right)\log(t)-L_2(t)
\label{f2}\end{equation}
is easily related to $L_2(t)$, using integration by parts.

{\bf Theorem 1}: Let
\begin{equation}
M_{k,n}(t)\equiv-\int_t^\frac1kL_{k-n-1}\left(\frac{x}{1-(n+1)x}\right)
\frac{f_n(x)\,{\rm d}x}{x(1-n x)}
\label{M}\end{equation}
for any pair of integers $k>n\ge0$. Then this integral evaluates to
\begin{equation}
M_{k,n}(t)=\sum_{m=0}^n(-1)^{n-m} L_{k-m}\left(\frac{t}{1-m t}\right)f_m(t).
\label{T1}\end{equation}

{\bf Proof}: At $n=0$, definition~(\ref{M}) gives $M_{k,0}(t)=L_k(t)$, by
virtue of recursion~(\ref{L}) for $L_k(t)$.
Hence~(\ref{T1}) holds at $n=0$. For $k>n+1>0$, recursion~(\ref{f})
for $f_{n+1}(t)$ allows an integration by parts in definition~(\ref{M}),
to obtain
\begin{equation}
M_{k,n}(t) = L_{k-n-1}\left(\frac{t}{1-(n+1)t}\right)f_{n+1}(t) - M_{k,n+1}(t)
\label{M1}\end{equation}
with a vanishing constant term, since the $L_{k-n-1}$ term vanishes at
$t=1/k$, where its argument $t/(1-(n+1)t)$ evaluates to $1/(k-n-1)$.
Hence I prove~(\ref{T1}) by induction, for all $k>n\ge0$.

{\bf Comment}: This is a very powerful result, peculiar to
the Dickman problem. For example, it allows us to
compute the Dickman heptalogarithm $L_7(t)$ very accurately,
as a single integral of a product of trilogarithms, instead of having
to evaluate a four-fold iterated integral of trilogs.

The key to the method is to observe that $f_{n+1}(t)=M_{n+1,n}(t)$.
To prove this, I set $k=n+1$ in the definition~(\ref{M})
of $M_{k,n}(t)$ and then use recursion~(\ref{f}) for $f_{n+1}(t)$.
I thus prove the claimed result~(\ref{f2}) for $f_2(t)$ by setting
$k=2$ and $n=1$ in the evaluation~(\ref{T1}) of Theorem 1.
Similarly, yet much more importantly, I obtain
\begin{equation}
f_3(t) = \log\left(\frac{t}{1-2t}\right)f_2(t)
-L_2\left(\frac{t}{1-t}\right)\log(t)+ L_3(t)
\label{f3}\end{equation}
by setting $k=3$ and $n=2$, thereby avoiding duplication
of the considerable effort expended in obtaining the
trilogarthmic result~(\ref{L3}) for $L_3(t)$.
Then, by setting $k=7$ and $n=3$ in the theorem,
I obtain the Dickman heptalogarithm
\begin{eqnarray}
L_7(t)&=&
 L_6\left(\frac{t}{1-t}\right)\log(t)
- L_5\left(\frac{t}{1-2t}\right)f_2(t)
+ L_4\left(\frac{t}{1-3t}\right)f_3(t)
\nonumber\\&&{}
+ \int_t^\frac17L_3\left(\frac{x}{1-4x}\right)
\frac{f_3(x)\,{\rm d}x}{x(1-3x)}
\label{f7}\end{eqnarray}
with a one-dimensional quadrature of products of known trilogs,
in the final term.
Moreover, there are two methods of evaluating $L_6(t)$
as an integral of products of known dilogs and trilogs,
using the pair $(L_2,f_3)$ or the pair $(L_3,f_2)$ in the final
integrand.
For $L_5(t)$ there are three methods, using $(L_3,f_1)$,
$(L_2,f_2)$ or $(L_1,f_3)$. Of these, the $(L_2,f_2)$ pair is the most
efficient. For $L_4(t)$ there are four methods, using
$(L_3,f_0)$, $(L_2,f_1)$, $(L_1,f_2)$ or $(L_0,f_3)$, where
$L_0(t)=f_0(t)=1$. Efficiency dictates that
one should use either the second or third pair; caution suggests
that one should use all four methods to check the accuracy
of numerical quadrature.

\section{Dickman heptalogs from standard trilogs}

To justify this methodology, I must first prove the claimed
result~(\ref{L3}), which reduces the Dickman trilog $L_3$ to
standard trilogs and establishes my claim that $C_3=-\frac13\zeta(3)$.
I begin by proving the rather simple identity
\begin{eqnarray}
L_3(t)&=&C_3 + C_2\log(t) + \frac16\log^3(t)
-{\rm Li}_3(t) + {\rm Li}_2(t)\log(t)\nonumber\\&&{}+
\int_0^t\left({\rm Li}_2\left(\frac{x}{1-x}\right)
+\frac12\,{\rm Li}_1^2(x)\right)\frac{{\rm d}x}{x}
\label{L31}\end{eqnarray}
where $C_2=-\frac{\pi^2}{12}$ was determined by~(\ref{L2}) and
$C_3$ is an integration constant, to be determined later by
the requirement that $L_3(\frac13)=0$. The proof of~(\ref{L31})
is symptomatic: we simply differentiate with respect to $t$
and check that $L_3^\prime(t)=L_2(t/(1-t))/t$, as required by~(\ref{L}).
Here, as ever, I use the relation
${\rm Li}_k^\prime(t)={\rm Li}_{k-1}(t)/t$,
with ${\rm Li}_1(t)=-\log(1-t)$.

Next comes a harder part,
namely to perform the integration
in~(\ref{L31}). Using integration by parts, I was able to
reduce the problem to an instance of equation 8.111 in the
fascinating and enormously informative book by Leonard Lewin~\cite{LL},
long since sadly out of print. Here,
I am content to state and then to prove that
\begin{eqnarray}
\frac74\,\zeta(3)&=&
\int_0^t\left({\rm Li}_2\left(\frac{x}{1-x}\right)
+\frac12\,{\rm Li}_1^2(x)\right)\frac{{\rm d}x}{x}
\nonumber\\&&{}
+ {\rm Li}_3(1-2t)
- {\rm Li}_3(2t-1)
+ {\rm Li}_3\left(\frac{-t}{1-2t}\right)
\nonumber\\&&{}
+ \left(-\,{\rm Li}_2(t)
+\frac16\,{\rm Li}_1^2(2t)
-\frac{1}{2}\,{\rm Li}_1(2t)\,{\rm Li}_1(1-t)
+\frac32\,{\rm Li_2}(1)
\right)\,{\rm Li}_1(2t)\,.
\label{L32}\end{eqnarray}
This is clearly true at $t=0$, since
${\rm Li}_3(1)=\zeta(3)$ and ${\rm Li}_3(-1)=-\frac34\zeta(3)$.
Thus it is sufficient to show that the right hand side of~(\ref{L32})
has a vanishing derivative. This derivative is of the form
$D_1(t)/t+D_2(t)/(1-2t)$, where $D_1$ and $D_2$ are rather
complicated combinations of dilogs and products of logs.
However, it is easy to show that
$D_1(0)=D_2(0)=0$. Hence, to prove~(\ref{L32}), it is
sufficient to show that $D_1^\prime(t)=D_2^\prime(t)=0$, which
may be done by elementary manipulation of logs.

Inverting the argument of ${\rm Li}_3(-t/(1-2t))$, in~(\ref{L32}),
and imposing the boundary condition $L_3(\frac13)=0$, in~(\ref{L31}),
I arrive at the claimed result~(\ref{L3}) for $L_3(t)$
and the claimed evaluation $3\,C_3=-\zeta(3)$, provided that
\begin{equation}3\left(2\,{\rm Li}_3\left(\frac13\right)
-{\rm Li}_3(-3)\right)-\log^3(3)=\frac{13}{2}\,\zeta(3).
\label{L33}\end{equation}

The final hurdle of proving~(\ref{L33}) was the most challenging.
Spencer Bloch and Herbert Gangl
told me that they expected the combination
$2\,{\rm Li}_3(\frac13)-{\rm Li}_3(-3)$ to evaluate
to some rational multiple of $\zeta(3)$, ``modulo logarithms".
Yet it appears that such rational numbers are as distressingly hard
to derive from first principles as they are disturbingly easy to guess
from low precision numerical computation.
My claim that $C_3=-\frac13\zeta(3)$ requires this rational
multiple to be $\frac{13}{6}$, with a denominator divisible by 3.
It was rather hard to see how a simple functional equation
for trilogs of a single variable might produce such
a denominator. Accordingly, I resorted to
the Spence--Kummer functional
relation for 9 trilogs of a pair of variables, given
in equation 6.107 of Lewin's book~\cite{LL}.
Setting $x = -1$ and $y = \frac13$
in that ornate identity and inverting the argument of
${\rm Li}_3(-\frac13)$, I proved~(\ref{L33}), obtaining
the combination $2\,{\rm Li}_3(1)-6\,{\rm Li}_3(-1)$ on the right hand
side. Thus the ``morally rational" coefficient of $\zeta(3)$
for the combination $2\,{\rm Li}_3(\frac13)-{\rm Li}_3(-3)$
is now proven to be $\frac13(2-6(-\frac34))=\frac{13}{6}$,
thereby rescuing this particular problem from what
Sasha Beilinson~\cite{Beil} memorably referred to as
``the burdock thicket of generalities".

Having thus proven the trilogarthmic input for the method
of Theorem~1, I am able to compute $F(\alpha)$ for $\alpha\ge\frac18$,
with great ease, and have provided 800 good digits for
\begin{eqnarray}
F(1/5)&\approx&
3.5472470045603972983389451077062356095164361057262
\times10^{-4}\label{F15}\\
F(1/6)&\approx&
1.9649696353955289651754986129204522894596719809623
\times10^{-5}\label{F16}\\
F(1/7)&\approx&
8.7456699532939166955802835727699721733804719764580
\times10^{-7}\label{F17}\\
F(1/8)&\approx&
3.2320693042261037725997853617282161576194751628024
\times10^{-8}\label{F18}
\end{eqnarray}
at the web page
{\tt http://physics.open.ac.uk/$\;\widetilde{}\;$dbroadhu/cert/smoctic.txt} .

A significant amount of LLL analysis produced no simple
integer relation of $F(1/5)$ to values of standards polylogs
and their products. It might be interesting to investigate
this issue more intensively, using David Bailey's
parallelization~\cite{BB} of Helaman Ferguson's
PSLQ~\cite{PSLQ} algorithm, since more than 800
digits of $L_4(1/5)$ may now be computed
from~(\ref{L}), using the explicit trilogarithms
in~(\ref{L3}). My own opinion, however, is that
the Dickman probability $F(1/5)$ does not reduce to standard polylogs.

\section{Dickman octalogs from Dickman tetralogs}

It was rather frustrating that Theorem~1
did not provide a method for computing Dickman octalogs
as single integrals of products of standard polylogs,
since it was precisely at weight 8 that I wished
to have high-quality numerical data with which to
determine whether the first MZV, namely $\zeta(5,3)$,
might show up in $C_8$. It was moral support from
Mike Oakes~\cite{MO} that determined me to push
the investigation above weight 7.

The barrier that must now be surmounted stems from the fact that
\begin{equation}
M_4(y)\equiv\int_0^y\left(
\log\left(\frac{x}{1+2x}\right)\,{\rm Li}_2(x)
+\frac12\log^2(x)\,{\rm Li}_1(-2x)
\right)\frac{{\rm d}x}{x(1+x)}
\label{M4}\end{equation}
appears in the Dickman tetralog,
since $L_4(t)-C_4+M_4(t/(1-2t))$ may be reduced to standard
polylogs and their products.
It is an easy matter to evaluate a single instance of~(\ref{M4}) to
high precision. For example, $M_4(\frac12)$ in~(\ref{C4}) was computed
to 800 digits, in order to determine the value of $C_4$
that results from the boundary condition $L_4(\frac14)=0$.
However, it is quite another matter to obtain good results for
$C_8$ and $C_9$, which are integrals with $M_4$ in their integrands.
When $t$ is significantly less than $\frac14$, one may efficiently
evaluate $M_4(t/(1-2t))$ by using Taylor series
for $\log(1+2x)$ and $1/(1+x)$
under the integral sign in~(\ref{M4}).
Yet to compute $C_8$ and $C_9$ we need $L_4(t)$
for all arguments $0<t<\frac14$,
inside the integrals for $L_8$ and $L_9$. So the remaining
problem involves an investigation of the behaviour of $L_4(t)$ in
the neighbourhood of $t=\frac14$.

{\bf Theorem 2}: Let $g_n(z)\equiv L_n(\frac1n-z)$ for $n>0$.
Then $g_n(z)$ has a Taylor series with rational coefficients,
beginning with $g_n(z)=G_n z^n+O(z^{n+1})$, where $G_n=
(-1)^n(n^n/n!)^2$.

{\bf Proof}: We know that $g_1(z)=\log(1-z)=-z+O(z^2)$
has a Taylor series with rational coefficients.
Setting $k=n+1$ in recursion~(\ref{L})
and transforming variables, I obtain
\begin{equation}g_{n+1}(z)=-\int_0^z
g_n\left(\frac{(n+1)^2y}{n^2+n(n+1)y}\right)\frac{(n+1){\rm d}y}{1-(n+1)y}.
\label{g}\end{equation}
Now suppose that the claim is true for $g_n$.
Then, by binomial expansion under the integral sign in~(\ref{g}),
it is also true for $g_{n+1}$, since
$G_{n+1}/G_n=-((n+1)^2/n^2)^n$. Hence, by induction, the claim
is true for all $n>0$.

{\bf Comment}: Starting with $g_1(z)=\log(1-z)$, I used three
iterations of recursion~(\ref{g}) to compute
sufficient terms in the rational Taylor series for
$g_4(z)=\frac{1024}{9}z^4+O(z^5)$ to achieve at least 240
digits of precision for $L_4(t)=g_4(\frac14-t)$ in the
region $\frac14>t>t_4=0.2358$. Then, for $t<t_4$,
I proceeded as follows.

Let $a_n=(-b_n+(-1)^n c_n)/n$, where
\begin{equation}
b_n=-\frac{3}{n^3}+\sum_{k=1}^n\frac{2^k{}k+(-1)^{k-1}n}{k^2n^2}
\label{b}\end{equation}
and $c_n$ is the coefficient of $z^n$ in the Taylor series for
$\log(1+2z){\rm Li}_2(z)/(1+z)$.
Then, with $x\equiv t/(1-t)$ and $y\equiv t/(1-2t)$, the
summation in
\begin{eqnarray}
L_4(t)-C_4&=&\sum_{n=2}^\infty\left\{a_n+b_n\log(y)\right\}(-y)^n
+L_3(x)\log(t) + {\rm Li}_2(-y)\log^2(y)
\nonumber\\&&{}
+\left({\rm Li}_2(-x)+\frac12\log^2(x)\right)
\left(\frac{\pi^2}{12}-\frac12\log^2(y)\right)
+\frac18\log^4(y)
\label{S4}\end{eqnarray}
provides a viable method for computing $L_4(t)$ for $t<t_4=0.2358$.

A sanity check was provided by comparing the
two strategies at $t=t_4$ and verifying that 240 good
digits of the conjectured value $\frac1{10}C_2^2$
were obtained for $C_4$. To achieve this, I expanded
$g_4(z)$ up to $O(z^{300})$ and took 4800 terms in the
summation over powers of $-y=-t/(1-2t)$ in~(\ref{S4}).
The relatively small number of terms used for $g_4$
resulted from the fact that it is rather laborious
to perform three rational binomial iterations of recursion~(\ref{g}).
By storing coefficients of the expansions, I am able to evaluate
240 good digits of the Dickman tetralog $L_4$ in less than 50
milliseconds and hence can efficiently compute the Dickman
octalog, $L_8$, and the nonalog, $L_9$, as one-dimensional quadratures
with integrands that call the procedure for $L_4$. By these means,
I arrived at conjecture~(\ref{C8}) for $C_8$ and then, thanks to
OEIS sequence A008991, inferred the wonderfully compact
conjectured generating function $\exp(\gamma z)/\Gamma(1-z)$
for $\sum_{k\ge0}C_k z^k$, which then gave
conjecture~(\ref{C9}) for $C_9$, also now verified at
high precision. Finally, I record 50 good digits of
the Dickman probabilities
\begin{eqnarray}
F(1/9)&\approx&
1.0162482827378365465348539356956957838244399586581
\times10^{-9}\label{F19}\\
F(1/10)&\approx&
2.7701718377259589887581212006343423263430066501156
\times10^{-11}\label{F110}
\end{eqnarray}
noting that 6 good digits were given in~\cite{LW},
which corrected serious errors in~\cite{BK}.

\section{Context and conclusion}

Thus far, I have spared the gentle reader explicit reference
to quantum field theory (QFT)
or to condensed matter physics.
I trust that s/he will now permit me to reveal the physics
context for investigating Dickman's
mathematical~\cite{Di,Ch,Rama,Br,DK,HT,BP,RB} problem.

The collision of a pair of protons in the
large hadron collider (LHC)~\cite{LHC},
at energies never before achieved in a particle accelerator, is
described by QFT in terms of the collision of quarks, or gluons,
carrying a fraction $x$ of the momentum of a proton.
The key thing that we need to know, to make sense of the
possible outcomes of these LHC collisions, is the probability
distribution function (PDF) in $x$. This is not known a priori; rather
it is inferred from electron-proton collisions at lower energies,
where half of the problem, namely the electron, is already
well understood. Yet the input PDF, from electron-proton collisions,
is not immediately usable at the LHC. Rather, it must be ``evolved",
up to the higher LHC energy. The procedure for doing this is
completely understood, in principle, yet involves formidable
mathematical challenges, in practice~\cite{BG,VVM}. It is performed as
a perturbation expansion in the coupling constant
of the strong interaction between quarks and gluons. At successive
orders in this compelling expansion, generalized
polylogarithms~\cite{RV,GR1}
in $x$ make their appearance.

From the perspective of QFT, the Dickman problem might
seem to be rather routine: as soon as the polylogarithmic
structure of the recursion~(\ref{L}) for $L_k$ is exposed,
it is clear that the formidable technical machinery
of QFT~\cite{RV,GR1,GR2,MUW,BBV,SSS,DDS,LSS}
has much to offer to the computation
of this problem in number theory. The fascinating circumstance
is that each iteration~(\ref{L})
of the Dickman polylog $L_k$ brings its own distinctive
constant $C_k$ in~(\ref{C}).
This also mimics QFT, where limiting values of generalized
polylogarithms are likewise zeta values, until one reaches
weight 8, when a
multiple zeta value appears~\cite{BK1,BGK,BK2,BBV}.
The Dickman problem would have
shared even more features with QFT had
$C_8$ contained this first irreducible MZV, namely
$\zeta(5,3)=\sum_{m>n>0} 1/(m^5n^3)$. In this paper I have
shown, with overwhelming probability, that $\zeta(5,3)$ does
not appear in $C_8$. Moreover, I have conjectured that $C_k$
is, most wonderfully, the coefficient of $z^k$ in the expansion of
$\exp(\gamma z)/\Gamma(1-z)$ and hence zeta-valued for all $k$.

This does not, however, mean that the Dickman problem now
lacks interest for physicists. On the contrary, it
resembles fascinating problems in condensed matter
physics. In the study of quantum spin chains~\cite{BVK1}, hugely demanding
calculations gave zeta-valued results up to weight 7. When
Dirk Kreimer and I asked Valdimir Korepin to climb the next
cliff, up to weight 8, we genuinely did not know whether an
MZV would appear. In fact it did not~\cite{BVK2}. Thus the Dickman
problem sits very well with the study of spin chains.

In conclusion, I suggest that the current investigation
has added to our understanding of the Dickman function
and has also confirmed how much more demanding is the
polylogarithmic structure of QFT, which appears to surpass
the Dickman function of number theory, and the spin chains of
condensed matter physics, both in its grand challenge and,
I believe, in its eventually to be comprehended great beauty~\cite{RPF}.

\subsection*{Acknowledgements}
This work was deeply influenced by my colleagues in physics,
Johannes Bl\"umlein,
Nigel Glover,
John Gracey,
Vladimir Korepin,
Dirk Kreimer,
Ettore Remiddi,
Volodya Smirnov and
Jos Vermaseren,
and by my colleagues in mathematics,
David Bailey,
Spencer Bloch,
Jonathan Borwein,
Helaman Ferguson,
Herbert Gangl
and Neil Sloane.
Yet, most of all, it derives from the
constant moral support of Mike Oakes.

\end{document}